# Microwave Photonic Ising Machine

Qizhuang Cen[1,†], Tengfei Hao[2,3,4,†], Hao Ding[1], Shanhong Guan[1], Zhiqiang Qin[1], Kun Xu[1], Yitang Dai[1,*], Ming Li[2,3,4,*].

[1]State Key Laboratory of Information Photonics and Optical Communications, Beijing University of Posts and Telecommunications, Beijing 100876, China.
[2]State Key Laboratory on Integrated Optoelectronics, Institute of Semiconductors, Chinese Academy of Sciences, Beijing 100083, China.
[3]School of Electronic, Electrical and Communication Engineering, University of Chinese Academy of Sciences, Beijing 100049, China.
[4]Center of Materials Science and Optoelectronics Engineering, University of Chinese Academy of Sciences, Beijing 100190, China.

†These authors contributed equally to this work.
*Corresponding authors: ytdai@bupt.edu.cn; ml@semi.ac.cn.

**Abstract:**
**Ising machines based on analog systems have the potential of acceleration in solving ubiquitous combinatorial optimization problems. Although some artificial spins to support large-scale Ising machine is reported, e.g. superconducting qubits in quantum annealers and short optical pulses in coherent Ising machines, the spin coherence is fragile due to the ultra-low equivalent temperature or optical phase sensitivity. In this paper, we propose to use short microwave pulses generated from an optoelectronic parametric oscillator as the spins to implement the Ising machine with large scale and also high coherence under room temperature. The proposed machine supports 10,000 spins, and the high coherence leads to accurate computation. Moreover, the Ising machine is highly compatible with high-speed electronic devices for programmability, paving a low-cost, accurate, and easy-to-implement way toward to solve real-world optimization problems.**

Combinatorial optimization problems can be found everywhere in modern society, such as drug discovery (*1,2*), finance (*3*), traffic flow optimization (*4*), and machine learning (*5*). Many combinatorial optimization problems are classified as the non-deterministic polynomial-time (NP)-hard or NP-complete complexity classes, which are hard to be solved on the standard digital computers since the number of combinations grows exponentially as the problem size $N$ increases. Although some software algorithms such as simulated annealing and other approximate algorithms (*6-9*) have been developed to accelerate the computation, using a Von Neumann architecture to solve those problems is still time-consuming due to the step-by-step computation mode and limited operating frequency (*10*). Interestingly, the solution to these problems can be effectively accelerated by mapping them onto an analog Ising machine. The Hamiltonian of a *N*-spin Ising machine without external field is given by $H = -\sum_{1\leq i<j\leq N} J_{ij}\sigma_i\sigma_j$, where $J_{ij}$ is the spin interaction between the *i*-th and $j$-th spins, $\sigma_i$ and $\sigma_j$ denote the z projection of the spins with eigenvalues of either 1 or −1. Finding the optimal answer is equivalent to searching the ground state of an Ising machine (*11*), and computing acceleration is obtained by utilizing the system intrinsic convergence property since convergence to the ground state is often at a high speed.

To solve the real-world optimization problems, large-scale, high-coherence Ising machines are highly desired (*1-5*). Many Ising machines, e.g. based on trapped ions (*12,13*), superconducting circuits (*14*), molecules (*15,16*), and optical systems (*17-21*) have been reported. The key features of these Ising spins are shown in Fig. 1. Although an Ising machine based on trapped ions has been successfully demonstrated with advantages such as high fidelity, long trapping and strong spin-spin couplings, it must operate under an ultra-low temperature of millikelvin and is constrained to small scale networks of about 300 spins due to the difficulties of constructing the requisite optical and electronic control systems (*13*). A large-scale Ising machine operates under room temperature can be achieved by using molecular spin, but it takes a long time of hundreds of seconds to optimize due to the limited speeds of molecular motions and chemical reactions (*16*). By the use of quantum superconducting or degenerate optical parametric oscillator (DOPO) spins, large-scale Ising machines have been implemented with computation acceleration. For example, the state-of-the-art quantum annealers based on superconducting circuits can achieve 5000 qubits/spins (*22*) and computation speedup can be realized thanks to the parallel search process (*23,24*). However, maintaining the spins coherence requires a temperature near absolute zero and extremely clean electromagnetic environment, which is costly and technically difficult. When spin values are represented by the phase of short optical pulses, Ising machines based on DOPOs can operate at room temperature, and more than 10,000 spins have been reported (*20*). Moreover, DOPO-based Ising machines can also search the answer parallelly (*25*) and is superior to quantum annealers (*26*) and digital computer (*27*) in dense graphic optimization. Nevertheless, as a long fiber cavity is used to store the large number of spins, it weakens the spin coherence since the optical phase is very sensitive to the cavity delay variation. According to $\Delta\varphi = 2\pi f_o \Delta t$, where $\Delta\varphi$ is the phase change of the spin, $f_o$ is the optical frequency (around 200 THz) and $\Delta t$ is the time jitter, a few femtoseconds of jitter would reverse the sign of the spin, causing the system Hamilton evolves away from the expectation. In short, the above-mentioned approaches have difficulties in implementing a long-term-operating, large-scale and high-coherence Ising machine under room temperature.

Here, we propose a novel Ising machine, whose artificial Ising spins are represented by the phases of the short microwave pulses generated from an optoelectronic parametric oscillator (OEPO) (*28*). Similar to the DOPO, the OEPO can also operate at degenerate state with relative phase of 0 or π, but with a much longer wavelength. Since the microwave wavelength (2 cm) is about 4 orders of that of the lightwave (1.5 μm), the microwave phase is much less sensitive to the cavity variation, resulting in high spin coherence. Moreover, by using low-loss, long optical fiber and time-division multiplexing technique, large scale spins can be obtained, proving enough space to code the real-world problems.

## Results
### Operation principle and experimental setup

Figure 2 shows the experimental setup and operation principle of the proposed microwave photonic Ising machine. One may figure out that the fabric of OEPO in Fig. 2A is similar to an optoelectronic oscillator (OEO) where a hybrid microwave photonics feedback loop has also been used. In essence, the conventional OEO is a continuous-wave microwave generator using a low-loss optical fiber as energy storage media to produce high-coherence microwave signals with an uncertain initial phase (*29-32*). In our scheme, the microwave phase is locked to a local oscillation (LO1) through electrical second-order degenerate parametric process. If the central frequency and the bandwidth of the microwave filter are properly designed, the generated microwave signal has a frequency half of LO1. In this case, the parametric process is a phase conjugate operator that reverse the phase of the input microwave signal (*28*). If the frequency of the LO1 is also carefully designed and synchronized to the cavity resonant frequency, the phase conjugate operator would lock oscillating frequency in a relative phase of either 0 or $\pi$ (see supplementary materials for details). The binary-phase oscillation is then used to simulate the artificial Ising spin, e.g. 0-phase/$\pi$-phase oscillation represents up/down spin. By using long fiber combined with time-division multiplexing, large-scale, discrete pulse oscillations are obtained in a single cavity. In such an oscillation network, the spin-spin interaction can be implemented by delay lines (*20*) or measurement-feedback circuit (*11*). The delay-line scheme, which is implemented by a wavelength-division multiplexing (WDM) system and optical delay lines (ODLs), is adopted in our demonstration. As shown in Fig. 2B, specific spin-spin interaction, defined by matrix *J*, would change the global loss of the OEPO network and result in particular phase configuration (see supplementary materials for details). Phase configuration with minimum loss has the maximum possibility to operate (*19*). By gradually increasing cavity gain, the OEPO network can operate around the minimum-loss state and outputs the corresponding microwave signal whose phase configuration is the optimal or suboptimal solution of the given Ising problem.

**Generation of 10,000 spins with high coherence**

Large scale and high coherence of the proposed system are verified in a non-interaction (*J* is a unit matrix) artificial spin network and results are shown in Fig. 3. The oscillation frequency of microwave photonic Ising spins is 10 GHz, and the wavelength is about 2 cm in the optical fiber with a refractive index of 1.47. The spin repetition period is 10 nanoseconds, so that 10,000 spins are supported by using 20-km fiber. Since the ratio between the cavity length and the oscillation wavelength is dramatically decreased from $10^9$ (*20*) to $10^6$, the spin values in the OEPO are much less sensitive to the inevitable temperature fluctuation and ambient vibration, obtaining a much higher coherence compared to the counterpart in DOPO.

Another external microwave source (LO2), whose frequency is the same as the oscillating signal and synchronized to the LO1, is used to demodulate the short microwave pulses to the baseband to extract the microwave phase. By tuning the phase of LO2, the demonstrated baseband pulses can reach the maximum peak-peak value, and the peak value in a single test (each test starts from noise to stable oscillation) is measured and shown in the histogram in Fig. 3A. The absolute value varies around a small range but may have opposite sign, which indicates the spins oscillate with stable amplitude and either 0 or $\pi$ phase. To further evaluate the spins coherence, 60,000 overlapped waveforms are recorded within 30 minutes by using a high-speed oscilloscope under persistence mode and results are shown in Fig. 3B. The "open eye" in Fig. 3B indicates that spins are highly stable both in amplitude and in phase, while the "eye" would be closed in an unstable system since its amplitude and/or phase varies over time. The high coherence guarantees the system evolves toward the expectation, thus accurate computation can be expected. This is a unique advantage over other Ising machines, such as based on DOPO, quantum superconducting circuits, or trapped ions, because maintaining high spin coherence is challenging in these networks. Since oscillation starts from shot noise and thermal noise, each spin is independent and global phase configuration would be chaotic if no interaction is implemented. In 100 tests, the ratio between the positive and negative spins of each test falls into the range from 0.97 to 1.02 (mean 0.9998), suggesting equal probabilities of 0-phase and π-phase oscillations. Furthermore, we extract the spin sign and calculate the correlation, as shown in Fig. 3C. In the autocorrelation, only one single peak is observed at the zero-bit delay, suggesting the spins are independent in each test. And no peak is observed along the whole span in the cross-correlation between two tests, which indicates that each test is independent.

**Simulation of 1D/2D Ising model**

Two Ising models, i.e., the closed one-dimensional (1D) chain and the 2D square lattice (*33*), are first used to verify the feasibility of the proposed Ising machine. To implement a 1D Ising simulator, a second channel with an additional 1-bit delay is used to interact the neighboring spins, as shown in Fig. 4A. The formed 1D Ising model is a closed loop and energy coupling is unidirectional ($J_{i,i+1} \neq 0, J_{i+1,i}=0$), from ($i-1$)-th spin to $i$-th spin for $i \leqslant 10,000$, and 10,000-th spin to the first spin. Figure 4B shows the evolutions of the spin phases and Ising energy as a function of roundtrip number in positive coupling 1D chain. The blue/yellow dot represents up/down spin. In our experiment, the cavity gain is gradually increased by enlarging the LO1 power. At the beginning, the cavity net gain is far below the oscillation threshold, so that system noise dominates the spin phases and leads to a chaotic configuration. As a result, the domains, namely different regions where spins have the same sign, are relatively short. As the LO1 power increases, cavity gain exceeds its loss, and the signal starts to oscillate along with the adjusting of the spin phases. From the noise to stable oscillation, some domains shrink and disappear, while others grow correspondingly. As the last domain wall (topological defect that separate different domains) falls down, all domains merges into one and all the spins have the same sign. This is because the positive energy couplings between spins forces the neighboring spins to share the same phase, so that the collaborative loss can be minimal. We calculate the frequencies of domains with different lengths as a function of the roundtrip number, and result is presented in Fig. 4C. In the chaotic stage, the short domains (domain length is below 10) have a relatively high frequency while the long domains are rare. As the roundtrip number increases, the spins are getting more and more orderly, and the frequencies of the short domain decrease. As the short domains collapse and merge into longer domains, the frequencies of the long domain increase in a certain period (roundtrip number from 2,000 to 4,000). As evolution continues, long domain continuously grows, swallowing up other short domains. The evolution stops at around 9,000 roundtrips with the spins oscillate stably in 0-phase state and the system reaches the ground state. Noted that the domain structure of the DOPO-based Ising machine with the same spin number would "freeze out" and will not reach the ground state unless one waits for an exponentially long time (*33*).

As shown in Fig. 5A, another channel with additional 100-bit delay is used to implement the vertical coupling. Combined with the above 1–bit horizontal coupling, we can simulate a 100×100 2D Ising square lattice. The ($i-1$)-th and ($i-100$)-th spins are coupled into the $i$-th spin with boundary conditions that the 10,000-th spin is coupled into the first spin, where $i \leqslant 10,000$, and the (9900+$k$)-th spin is coupled into the $k$-th spin, where $1 \leqslant k \leqslant 100$. Thanks to the additional constraints brought by the interaction of the vertical dimension, the machine takes much less time to evolve from the chaotic state to the ordered state compared to the 1D Ising simulator, as shown in Fig. 5C. A much steeper curve can also be observed in the Ising energy evolution. Only tens of roundtrips are taken to reach a low energy close to the ground state, suggesting a clear phase transition between the chaos and the order state. Some snapshots at specific roundtrips are shown in Fig. 5B to reveal the evolution process. Here, the coupling coefficients in two dimensions are both positive and unidirectional. As a result, only one domain remains and all spins fall into the same phase configuration at the end. The frequencies of domains with different sizes as a function of roundtrip number are shown in Fig. 5D. As the roundtrip number increases, the large domain gradually swallows up other domains and the machine takes less than 1,200 roundtrips to the ground state.

**Solving a Max-cut problem**

Max-cut problem is an NP-hard problem that can be mapped onto Ising formulation (*11*). Here, we program an unweighted, unidirectional Möbius ladder graph with 20 vertices into our machine, as shown in insect of Fig. 6A. In this demonstration, the 20-km fiber is replaced by a short one, and the 100-bit delay line is also replaced by a 10-bit one while other parameters are maintained. In a graph with 20 vertices, the number of possible combinations is $2^{20}$. Although the possible Ising energy has only 27 different values which vary from −26 to 30, finding the lowest states is still very difficult since there are only twenty out of $2^{20}$ combinations meet the requirement, as shown in Fig. 6A. By tuning the according ODLs, the desired interaction matrix is implemented. Figure 6B shows the demodulated baseband pulses train. The Ising energy is −26 when the oscillation is stable, suggesting the lowest energy under the given interaction matrix is found. We performed 100 tests and find that lowest Ising energy is obtained in each test. The minimum Ising energy corresponds to the max cut, which is 28. Figure 6C and D present each spin evolutions and the Ising energy/graph cut size as a function of round-trip number. The spins have random phase at the beginning, corresponding to the relatively high Ising energy and small cut. As the round-trip number increases, spins evolve towards certain, either 0 or π phase. Meanwhile, the Ising energy drops down while the cut size increase. At around 60 roundtrips, the machine reaches the lowest-energy phase configuration and the given Max-cut problem is solved.

**Discussion**

In order to solve combinatorial optimization problems in real-world applications, the Ising machine should be qualified with performances such as large scale, high coherence, and also flexible programmability. Large-scale spin network provides enough space for coding real-world complex problems. High coherence ensures accurate computation of the Ising machine. Programmability makes the machine be common to use for various optimization problems. Thanks to the use of low-loss, long optical fiber in the optoelectronic cavity of OEPO, a large-scale Ising network can be easily implemented. Since the spins in the proposed machine are represented by short microwave pulses, which are less sensitive to the environment disturbance and maintain high coherence, the machine evolves fast toward the ground state. Moreover, flexible programmability can also be easily realized by utilizing measurement-feedback schemes, since the microwave spin can be easily measured and controlled with high-speed electrical devices (*21*). In a word, the microwave photonic Ising machine promises an optimizer with a large scale, high coherence, and programmability, paving a way to solve real-world problems.

**Methods**
**Experimental details**

A multi-channel laser (IDPhotonics, CoBriteDX4) is used in our experiment. In the non-interaction oscillation, only one channel is used. In the simulation of 1D Ising model, two channels are used. In 2D Ising model simulation and solving the Max-cut problem, three channels are used. These channels have a 100-GHz frequency spacing with the wavelength from 1549.32 nm to 1550.92 nm. The continuous lightwave from the laser are coupled together through a WDM and then shaped into optical pulse trains by using electrical pulse trains and a low-biased intensity modulator. The electrical pulse trains are generated from a programmable pulse pattern generator (PPG, Keysight N4960A) with a repetition rate of 100 MHz and a duty cycle of 20%. The generated optical pulse trains are then used as the carrier of the oscillating microwave signal in the optoelectronic cavity. A multiplexer and a demultiplexer are used in the cavity to enable independent delay control for each channel. Three tunable ODLs are inserted between the multiplexer and the demultiplexer. The tunable range of the ODLs is 500 ps, which is much larger than the period of the oscillating microwave signal (100 ps). As a result, the sign of coupling coefficients in interaction matrix *J* can be easily adjusted, while their magnitudes are also alterable by tuning the laser power. To minimize the dispersion, a 20-km DSF is used to store the microwave spins. The PD (DSC40S) in our experiment has 3-dB bandwidth of 18 GHz and responsivity of 0.75 A/W at 1550 nm. The signal after PD is filtered by an 8-12 GHz BPF and amplified by an electrical amplifier, then is divided into two parts: one is used for measurement and another one is launched into an electrical mixer and interacts with an external oscillation (LO1) with 20-GHz frequency and 10-dBm power. A second 8-12 GHz BPF is used to suppress the sum-frequency signal and leaked LO1. The filtered IF signal is then amplified and feedback to a quadrature-biased intensity modulator (IM). The LO1 frequency $f_{LO1}$, the carrier frequency of the microwave pulses $f_s$, and the free spectrum range (*FSR*) of the optoelectronic cavity $f_{FSR}$ satisfy $f_{LO1} = 2f_s = 2Mf_{FSR}$, where *M* is an integer.

The microwave pulses (Ising spins) oscillate from noise in the OEPO with random phase, then iterate in the cavity with increasing amplitude and their phases converge to either 0 or π state. Since the signal bandwidth is about 1 GHz, which is much less than the channel frequency spacing, the lightwave interference between different channels has no contribution to the microwave pulses. Another external microwave source (LO2), which is synchronized to LO1, with a frequency of $f_{LO1}/2$ and 10-dBm power is used to extract the phases of the microwave pulse. The demonstrated baseband signal is a short microwave pulse train with positive or negative amplitude. The signal passes through a low-pass filter (LPF) with 3-dB bandwidth of 1 GHz and then launched into a high-speed, real-time oscilloscope (Tektronix, DPO70000) and an extra digitizer (ADLINK PXIe-9852). The digitizer with an analog bandwidth of 90 MHz and 14-bit quantization is synchronized to the PPG and samples the demodulated baseband signal with 100-MHz frequency. In this case, only a single point is sampled for each baseband pulse. By using a tunable electrical delay line, one can ensure that the sampling is just at the peak of the pulse.

**References**

1. Kitchen, D. B., Decornez, H., Furr & J. R., Bajorath, J. Docking and scoring invirtual screening for drug discovery: methods and applications. *Nature Rev. Drug Discov.* **3**, 935–949 (2004).
2. Gordon, E. M., Gallop, M. A. & Patel, D. V. Strategy and tactics in combinatorial organic synthesis. Applications to drug discovery. *Acc. Chem. Res.* **29**, 144–154 (1996).
3. Hoesel, S. van & Müller, R. Optimization in electronic markets: examples in combinatorial auctions. *Netnomics*, **3**, 23–33 (2001).
4. Neukart, F. et al. Traffic flow optimization using a quantum annealer. *Front. ICT* **4**, 29 (2017).
5. Bojnordi, M. N. & Ipek, E. Memristive Boltzmann machine: a hardware accelerator for

combinatorial optimization and deep learning. In *2016 IEEE International Symposium on High Performance Computer Architecture (HPCA)* 1–13 (2016).

6. Kirkpatrick, S., Gelatt Jr., D. & Vecchi, M. P. Optimization by simulated annealing. *Science* **220**, 4598 (1983).
7. Papadimitriou, C. H. & Steiglitz, K. Combinatorial optimization: algorithms and complexity (Courier Dover Publications, 1998).
8. Hochbaum, D. S. Approximation algorithms for NP-hard problems (PWS Publishing Co., 1996)
9. Goemans, M. X. & Williams, D. P. Improved approximation algorithms for maximum cut and satisfiability problems using semidefinite programming. *J. of the ACM* **42**, 1115 (1995).
10. Waldrop, M. M. The chips are down for Moore's law. *Nature* **530**, 144–147 (2016).
11. McMahon, P. L. et al. A fully programmable 100-spin coherent Ising machine with all-to-all connections. *Science* **354**, 614–617 (2016).
12. Kim, K. et al. Quantum simulation of frustrated Ising spins with trapped ions. *Nature* **465**, 590–593 (2010).
13. Bruzewicz, C. D.. Chiaverini, J., McConnell, R. & Sage, J. M. Trapped-ion quantum computing: Progress and challenges. *Applied Physics Reviews* **6**, 021314 (2019).
14. Johnson, M. W. et al. Quantum annealing with manufactured spins. *Nature* **473**, 194-198 (2011).
15. Adleman, L. M. Molecular computation of solutions to combinatorial problems. *Science* **266**, 1021-1024 (1994).
16. Guo, S. et al. A molecular computing approach to solving optimization problems via programmable microdroplet arrays. *ChemRxiv. Preprint.* https://doi.org/10.26434/chemrxiv.10250897.v1 (2019).
17. Oltean, M. Solving the Hamiltonian path problem with a light-based computer. *Nat. Comput.* **7**, 57–70 (2008).
18. Utsunomiya, S., Takata, K. & Yamamoto, Y. Mapping of Ising models onto injection-locked laser systems. *Opt. Express* **19**, 18091–18108 (2011).
19. Wang, Z., Marandi, A., Wen, K., Byer, R. L. & Yamamoto, Y. Coherent Ising machine based on degenerate optical parametric oscillators. *Phys. Rev. A* **88**, 063853 (2013).
20. Marandi, A., Wang, Z., Takata, K., L. Byer, R. & Yamamoto, Y. Network of time-multiplexed optical parametric oscillators as a coherent Ising machine. *Nat. Photonics* **8**, 937–942 (2014).
21. Lucas, A. Ising formulations of many NP problems. *Front. Phys.* **2**, 5 (2014).
22. Russell, J. D-Wave previews next-gen platform; debuts Pegasus topology; targets 5000 qubits. *HPCwire*, https://www.hpcwire.com/2019/02/27/d-wave-previews-next-gen-platform-debuts-pegasus-topology-targets-5000-qubits/ (2019).
23. Waidyasooriya, H. M., & Hariyama, M. Highly-parallel FPGA accelerator for simulated quantum annealing, *IEEE Trans. Emerg. Topics Comput.* in press (2020).
24. Waidyasooriya, H. M., & Hariyama, M. A GPU-Based Quantum Annealing Simulator for Fully-Connected Ising Models Utilizing Spatial and Temporal Parallelism. *IEEE Access* **8**, 67929-67939 (2020).
25. Yamamoto, Y. et al. Coherent Ising machines—Optical neural networks operating at the quantum limit. *npj Quantum Inf.* **3**, 49 (2017).
26. Hamerly, R. et al. Experimental investigation of performance differences between Coherent Ising machines and a quantum annealer. *Sci. Adv.* **5**, eaau0823 (2019).
27. Inagaki, T. et al. A coherent Ising machine for 2000-node optimization problems. *Science* **354**, 603–606 (2016).

28. Hao, T. et al. Optoelectronic parametric oscillator. *Light Sci. Appl.* **9**, 102 (2020).
29. Maleki, L. Optoelectronic oscillators for microwave and mm-wave generation. in *Proc. 18th Int. Radar Symp.* 1-5, (2017).
30. Yao, X. S. & Maleki, L. Optoelectronic microwave oscillator. *J. Opt. Soc. Am. B* **13**, 1725–1735 (1996).
31. Hao, T. et al. Breaking the limitation of mode building time in an optoelectronic oscillator. *Nature Commun.* **9**, 1839 (2018).
32. Levy, E. C., Horowitz, M. & Menyuk, C. R. Modeling optoelectronic oscillators. *J. Opt. Soc. Am. B* **26**, 148–159 (2009).
33. Hamerly, R. et al. Topological defect formation in 1D and 2D spin chains realized by network of optical parametric oscillators. *Int. J. Mod. Phys. B* **30**, 1630014 (2016).

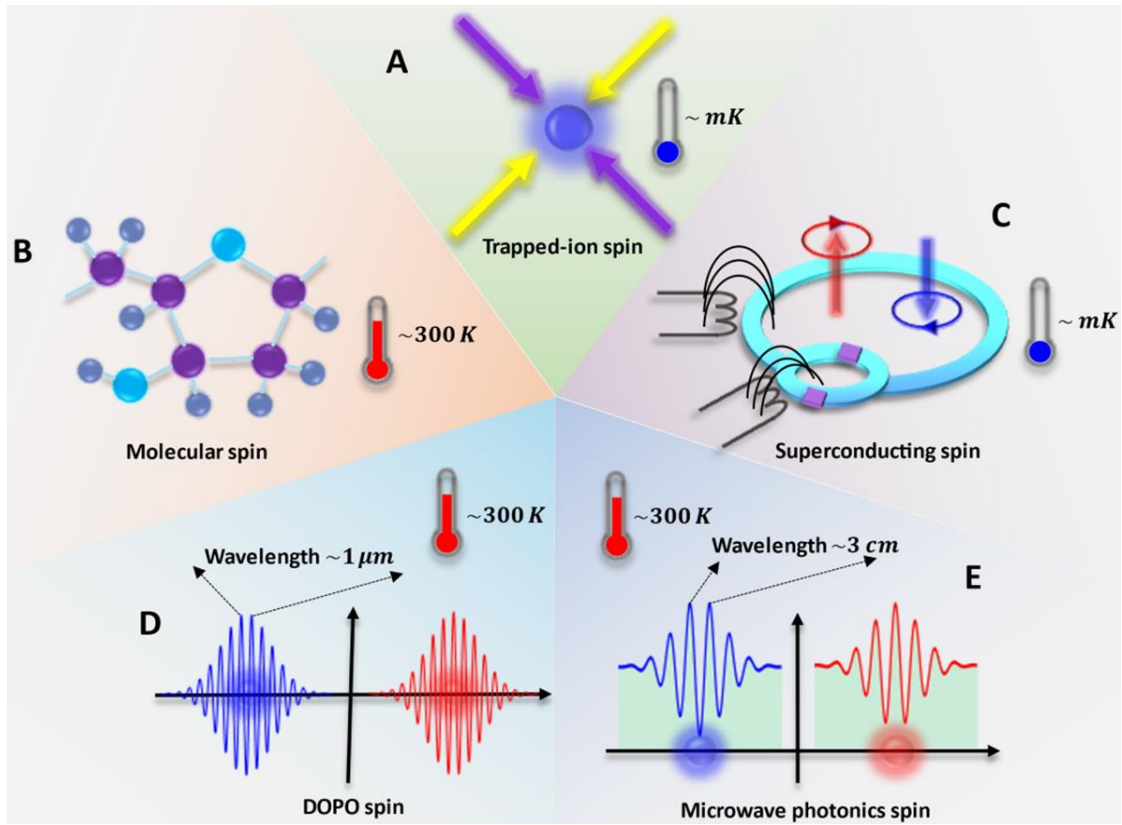

**Fig. 1. Comparisons among different Ising spins.** **(A)** Trapped-ion spin, operates at millikelvin, is hard to implement large-scale network and also costly. **(B)** Molecular spin, operates under room temperature, is easy to realize large-scale network but takes a long time to search the solution. **(C)** Superconducting spin, operates at millikelvin, has a prospect of large scale but costly. **(D)** DOPO spin, operates under room temperature, is easy to implement large-scale network, but needs precise control of the cavity length to maintain spin coherence. **(E)** The proposed microwave photonic spin, operates under room temperature, is easy to implement large-scale and high-coherence network with low cost.

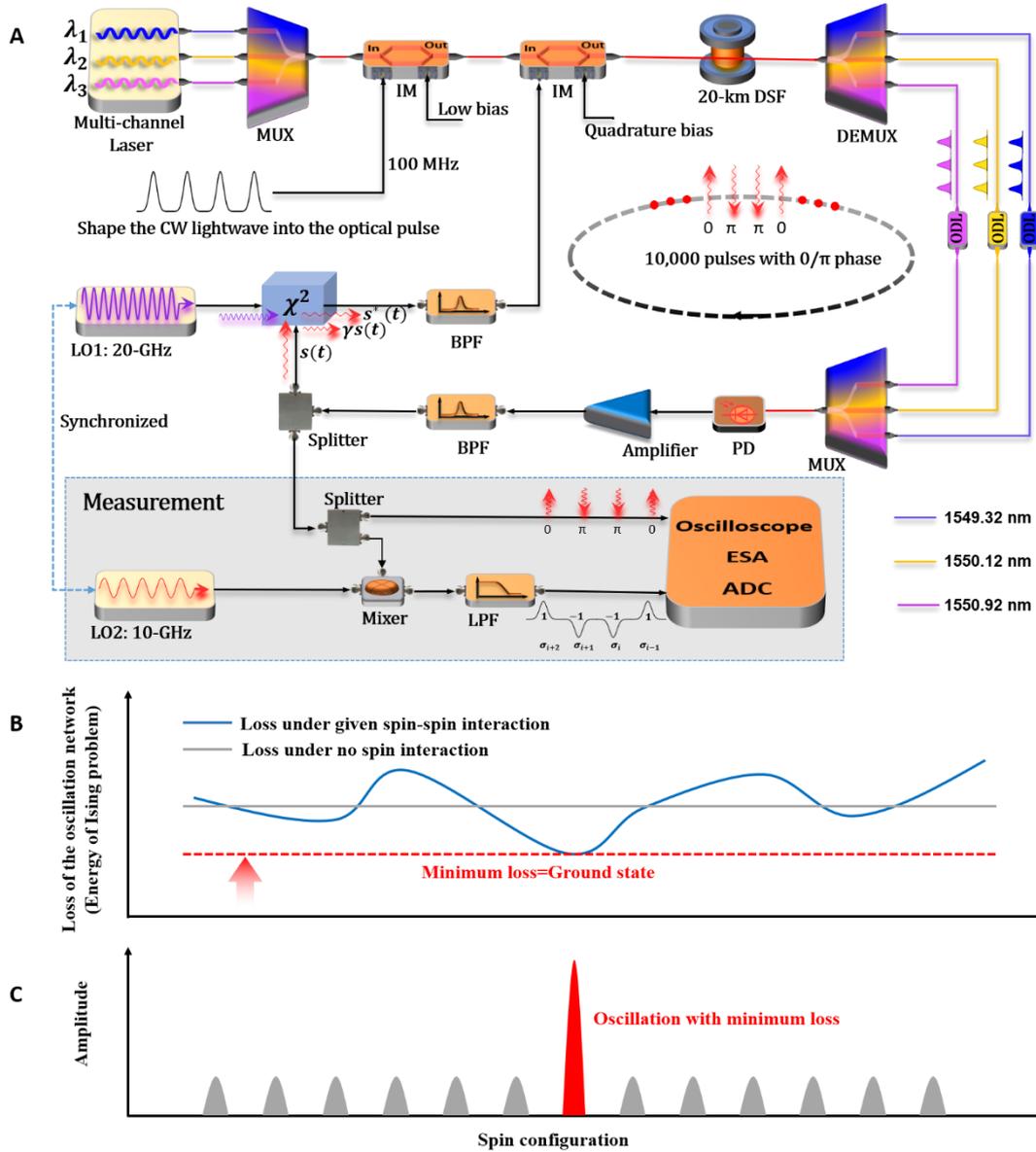

**Fig. 2. Schematic diagram and operation principle of the proposed Ising machine based on an OEPO. (A)** Experimental setup. Spins are represented with binary phases of short microwave pulses that are generated from OEPO and spin interactions are realized by using different channels in the optical path with configurable delays; MUX: multiplexer; DEMUX: demultiplexer; IM: intensity modulator; DSF: dispersion-shifted fiber; ODL: optical delay line; PD: photodetector; BPF: bandpass filter; LPF: low-pass filter; LO: local oscillation (LO1 with 20-GHz frequency is used to lock the microwave phase either in 0 or π, LO2 with 10-GHz frequency is used to demodulated the oscillating microwave phase); ESA: electrical spectrum analyzer; ADC: analog to digital converter. (**B**) A given spin-spin interaction would change the network global loss and the minimum-loss operation can be achieved by gradually increasing the gain of the OEPO network. (**C**) The network with minimum-loss spin configuration has the maximum possibility to oscillate, and the phase configuration corresponds to the given Ising problem.

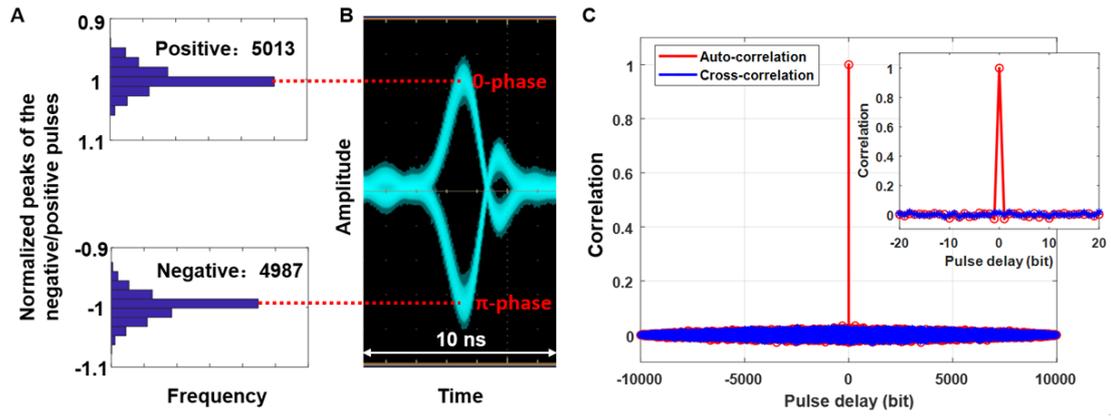

**Fig. 3. Experiment results of non-interaction oscillation.** (**A**) Normalized peaks of the demodulated baseband pulses. Pulses with positive/negative amplitude have a phase of 0/π. (**B**) Overlay of 60,000 traces in 30 minutes. (**C**) Auto-correlation in a test and cross-correlation between two tests.

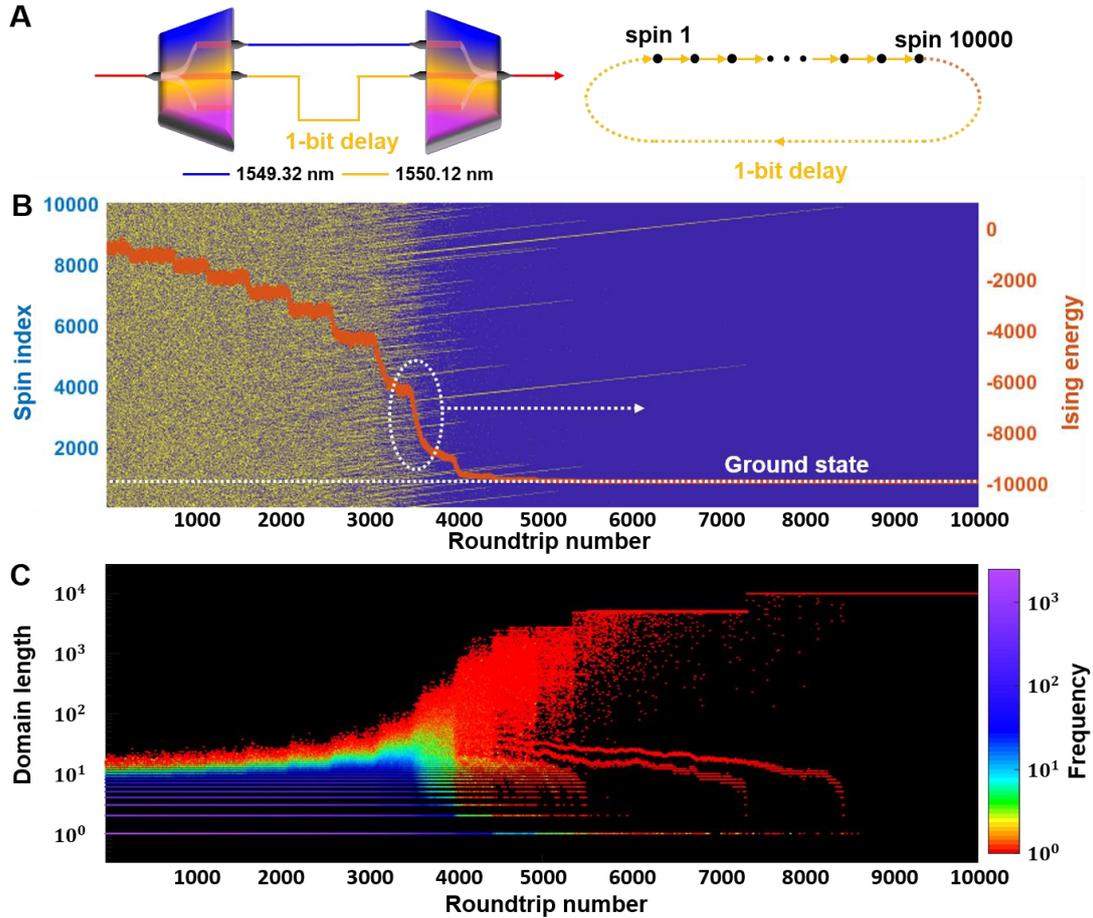

**Fig. 4. Experimental results of 1D Ising simulator.** (**A**) Schematic of the 1D Ising simulator. Left: two-channel WDM system with 1-bit delay. Right: 1D Ising graph. (**B**) Evolutions of the spin phases and Ising energy as a function of round-trip number. The blue/yellow dot represents spin with 0/π phase. By gradually increasing the cavity gain, spin values are adjusted with the increases of the roundtrip number. Finally, all the spins share the same value since the energy couplings are positive. (**C**) Frequencies of specific length domain as a function of roundtrip number. The domains are different regions where spins have the same value. The frequencies of the short domains decreases with the increases of roundtrip number and the Ising machine reaches the ground state at around 9,000 roundtrips in which there is only one domain with a length of 10,000.

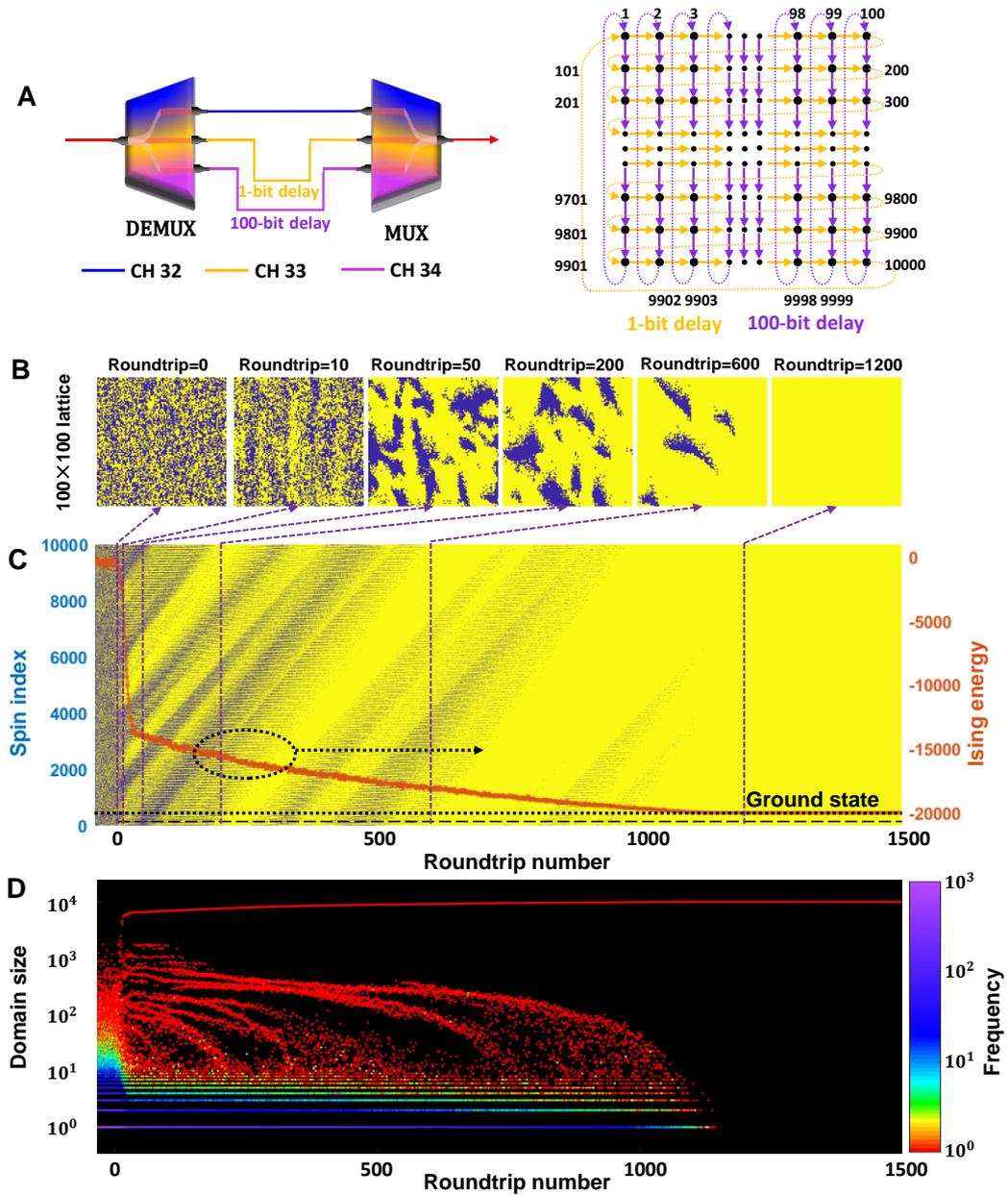

**Fig. 5. Experimental results of 2D Ising simulator.** (**A**) Schematic of the 2D Ising simulator. Left: three-channel WDM system with 1-bit delay and 100-bit delay. Right: 2D Ising graph. (**B**) Snapshots of the spin evolution at specific roundtrip. (**C**) Spin evolution and Ising energy as a function of round-trip number. The machine takes much less time to transition from a chaotic state to an ordered state compared to that in the 1D Ising simulator owning to the added constraints brought by the interaction of an extra dimension. (**D**) The frequencies of specific size domain as a function of the roundtrip number.

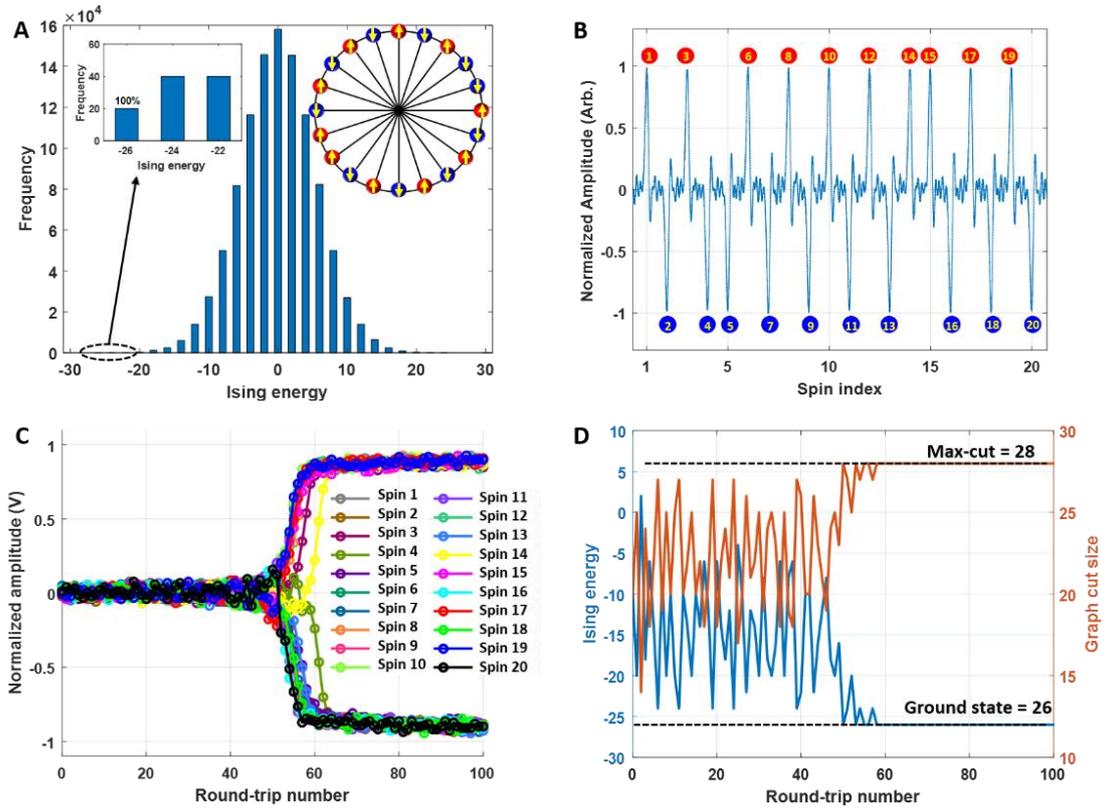

**Fig. 6. Results with N=20 cubic graphs. (A)** A Möbius ladder graph with N=20 vertices, and Ising energy frequencies of $2^{20}$ possible phase configurations. **(B)** Demodulated baseband pulses. **(C)** The spin evolutions as a function of the round-trip number. **(D)** The evolution of the graph cut size and Ising energy as a function of the round-trip number.

# Supplementary Materials for

# Microwave Photonic Ising Machine

1. **Binary-phase oscillation in degenerate OEPO**

In a conventional optoelectronic oscillator (OEO) cavity, microwave signal is up-converted to optical domain through an electro-optic modulator, and then recovers itself in a photodetector (PD) after propagates along a long optical fiber. The recovered microwave signal is then amplified, filtered, and finally feedback to the modulator. In an optoelectronic parametric oscillator (OEPO), microwave signal passes through the optoelectronic cavity and interacts with a local oscillation (LO) through an electrical second-order nonlinear device, i.e. an electrical mixer. The second-order parametric process produces the sum frequency and the difference frequency of the two input signals. In our scheme, the difference frequency is reserved and serviced as the artificial Ising spin while the sum frequency is blocked by a microwave bandpass filter. If the central frequency and the bandwidth of the microwave filter are properly designed, the signals before and after the frequency conversion, known as radio frequency (RF) and intermediate frequency (IF) signal respectively, would have the same frequency. In this case, the parametric process is degenerate and there would be a phase conjugate operation in the mixer that reverses the phase of the input RF signal (*28*). The phase-reversed RF signal transfers to the IF port and mixes with the leaked RF signal, which directly transfers to the IF port without interaction with the LO signal. In a commercial electrical mixer, a typical rejection between the leaked RF and IF (phase-reversed RF signal) is from −10 dB to −30 dB. Since the leaked RF and the phase-reversed RF signal have the same frequency, they cannot be separated from each other. In this case, the loss of the signal passing through the mixer depends on the phase difference between the phase-reversed one and the leaked one. The minimum loss is realized when the two parts have the same phase. As a result, the signal before and after the parametric process are the same, which means phase of the oscillating microwave signal is either 0 or π. The binary-phase oscillation then can be used to represent the artificial spin of the Ising machine. The spin evolution in the proposed optoelectronic cavity can be described as (*28, 32*)

$$s(t+\tau) = \alpha \left\{ \sqrt{G_{EA}} R_{PD} I_{PD} Z_{PD} J_1\left(\frac{\pi |s(t)|}{V_\pi}\right) e^{i \arg[s(t)]} + n(t) \right\}^* \cdot p(t) + \gamma s(t), \quad (S1)$$

where $s(t)$ is the oscillating signal, $\tau$ is the cavity delay, $\alpha$ is the frequency conversion coefficient of the mixer, $G_{EA}$ is the power gain of the electrical amplifier, $R_{PD}$ is the responsivity of the PD, $I_{PD}$ is the power that launched into the PD, $Z_{PD}$ is the PD impedance, $J_1$ is the 1st-order Bessel function of the first kind, $V_\pi$ is the half-wave voltage of the modulator, $p(t) = Ae^{i\omega_{LO}t}$ is the external microwave signal with fixed amplitude $A$ and frequency $\omega_{LO}$, $\gamma$ is the signal leakage coefficient of the mixer, and $n(t)$ is the link noise which includes thermal noise and shot noise. In the OEPO, oscillation starts from noise and the noise iterates in the cavity after each roundtrip and obtains constructive or destructive interference. With enough cavity gain, the interference finally gets stable and oscillator outputs the desired microwave signal.

Here, we run a simulation with Matlab. The simulation is run at baseband and parameters are modeled as follows: $\alpha = 0.4$, $G_{EA} = 40\ dB$, $R_{PD} = 1\ A/W$, $I_{PD} = 6\ dBm$, $Z_{PD} = 50\ \Omega$, $V_\pi =$ $2\pi$ volts, $p(t) = 1$, $\gamma = 0.1$ and the noise floor is modeled as Gaussian white noise with power spectral density of −150 dBm/Hz. The frequency conversion coefficient $\alpha = 0.4$ corresponds to a power loss of 8 dB in the mixer. The leakage coefficient $\gamma = 0.1$ corresponds to a power rejection of 20 dB between the leaked RF and IF. The cavity net gain is calculated to be 2 dB. The results of ten spins are shown in Fig. S1. After less than 30 roundtrips, the oscillation amplitude reaches the maximum and then stays stable, as can be seen in Fig.1SA. Different from the monotonic increasing in the amplitude, the spin phase jumps back and forth at the beginning because of the phase conjugate operator, as shown in Fig. S1B. As the roundtrip number increase, the wobbles become smaller and finally the spin phases converge to either 0 or π. The stability of the oscillation is also evaluated by the calculating the jitter $\frac{|s(t+\tau) - s(t)|}{|s(t)|}$ and results are shown in Fig. S1C. One can conclude that the spin is highly stable and jitters drops down to around $10^{-6}$ after 100 roundtrips. The cavity net gains are around 2 dB at the beginning and decrease to one when the spin amplitudes are stable.

## 2. Mapping the Ising problem onto the OEPO network

In an *N*-spin, non-interaction oscillation network, assuming each of spin has a normalized loss of one, we obtain the global loss *N*. If the *i*-th spin is injected by the *j*-th spin with a coupling coefficient of $J_{i,j}$, the loss of the *i*-th spin would be changed. If the sign of $J_{i,j}\sigma_i\sigma_j$ is positive/negative, the loss would decrease/increase from 1 to $1 - J_{i,j}\sigma_i\sigma_j$. As a result, all the couplings from other spins have an integrated loss contribution of $-\sum_{1 \leq j \leq N, j \neq i} J_{ij}\sigma_i\sigma_j$, and the total loss of the *i*-th spin is then given by $\Gamma_i = 1 - \sum_{1 \leq j \leq N, j \neq i} J_{ij}\sigma_i\sigma_j$ (*19*). Accordingly, the global loss of the *N*-spin oscillation network can be expressed as

$$\Gamma = \sum_{1 \leq i \leq N} \Gamma_i = N - \sum_{1 \leq i, j \leq N} J_{ij}\sigma_i\sigma_j. \qquad (S2)$$

Noted that $-\sum_{1 \leq i, j \leq N} J_{ij}\sigma_i\sigma_j$ is the Hamiltonian of the *N*-spin Ising model without external field, so that we can conclude that the potential phase configurations in an *N*-spin oscillation network under the given interaction matrix ***J*** corresponds to the global loss and maps onto the Ising energy landscape of a given Ising model. Based on the minimum-loss principle, the oscillation network most likely operates at this state. To solve an Ising problem, one can program the corresponding matrix ***J*** into the oscillation network, and then gradually increase the cavity gain to search the minimum-loss state. As the cavity gain exceeds its loss, oscillation starts with an increase in amplitude and adjustment of phase. When the oscillation network runs stably, one can measure the corresponding phase configuration, from which the given problems can be solved.

## 3. Other experimental results

The temporal waveforms of the oscillating microwave pulses and the demodulated baseband pulses are shown in Fig. S2A. The microwave pulse lasts several periods. In the non-interaction oscillation, 100 tests are performed. The ratios between the numbers of the positive and negative pulses are calculated and the results fall into the range from 0.97 to 1.02, as shown in Fig. S2B. The power spectrum of the non-interaction oscillation is also measured by an electrical spectrum analyzer (ESA, Keysight N9030A) and the results are shown in Fig. S3C. One can find that the power spectrum is actually a microwave frequency comb whose frequency spacing equals to the FSR of the optoelectronic cavity. Clear comb line under high-resolution bandwidth (RBW) suggests the microwave signal is highly stable, which corresponds to high spin coherence.

The temporal waveform of the demodulated baseband pulses in the 1D positive coupling Ising simulation is shown in Fig. S3A. Since the energy couplings are positive, all the spins share the same value. The period of the signal is 10 ns, which equals to the interval of two neighboring spins. This results in a much sparser comb line in the frequency domain, as shown in Fig. S3B and C.

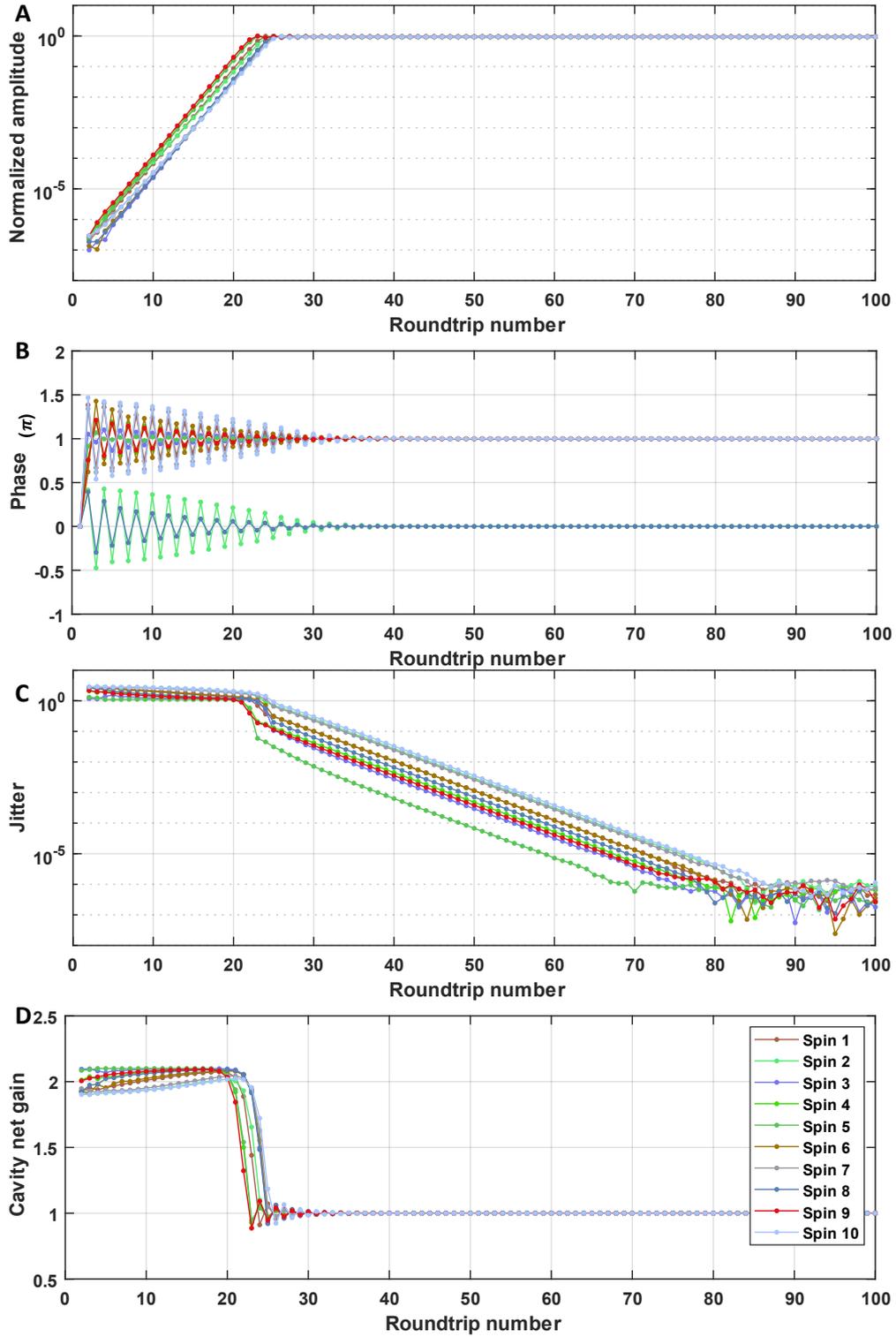

**Fig. S1. Simulation of the Ising spin of the proposed microwave photonics Ising machine.** (**A**) Spin amplitude evolution. (**B**) Spin phase evolution. (**C**) Jitter of the spin as a function of round-trip number. (**D**) Cavity net gain as a function of roundtrip number.

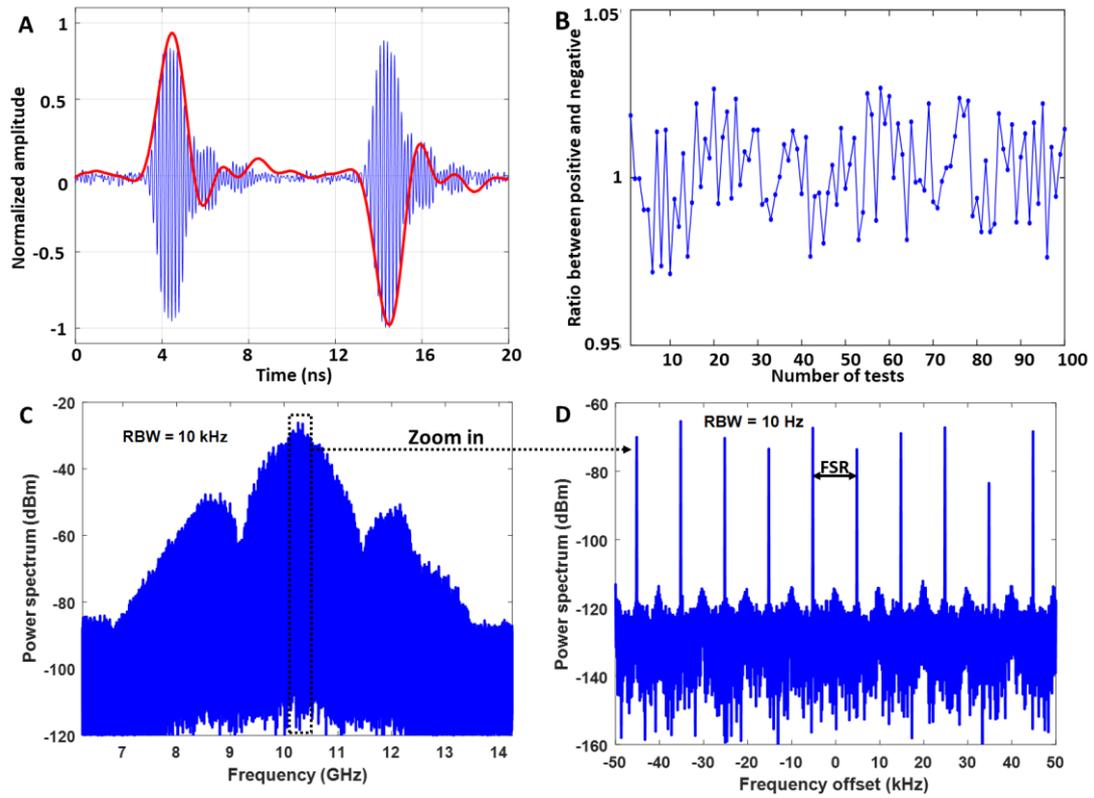

**Fig. S2. Experimental results of non-interaction oscillation.** (**A**) The temporal waveforms of the short microwave pulses and the demodulated baseband pulses; (**B**) The ratios between the positive and negative pulses in 100 non-interaction tests; (**C**) 6-GHz span power spectrum in 100-kHz resolution bandwidth (RBW); (**D**) Detail of power spectrum in 10-Hz RBW.

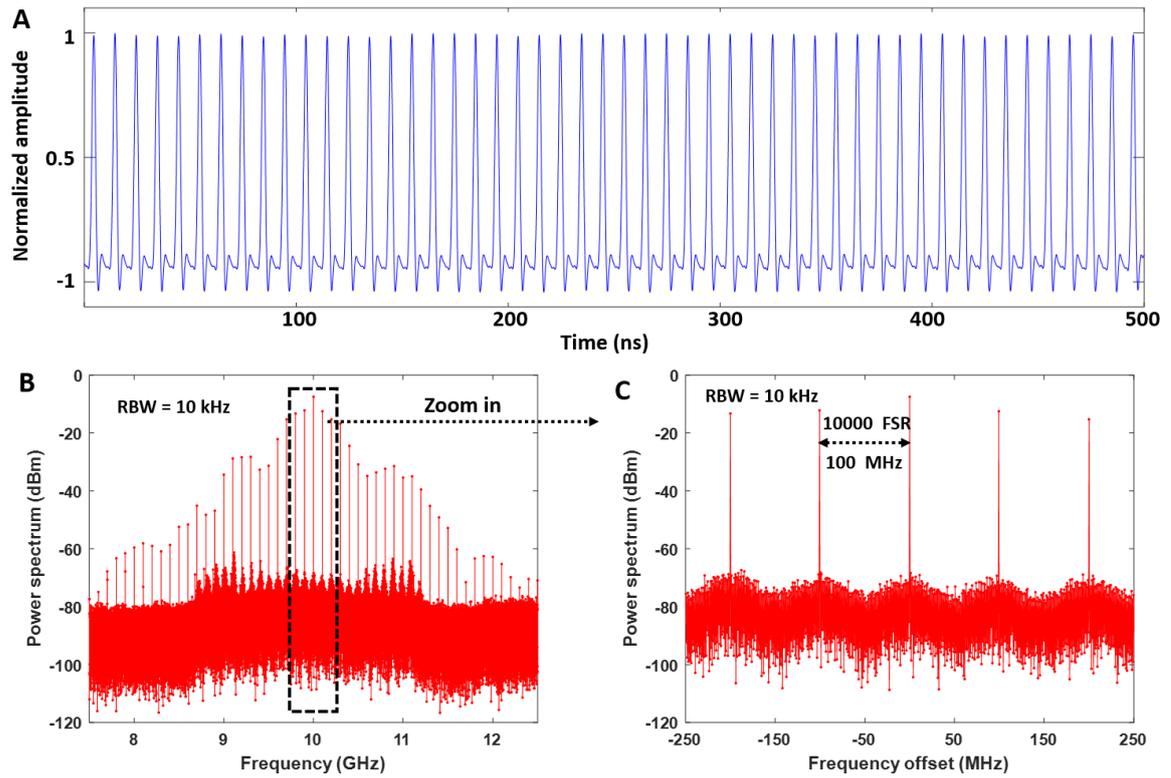

**Fig. S3. Temporal waveform and spectrum of the 1D Ising simulator.** (**A**) The temporal waveform of the demodulated baseband pulse; (**B**) 5-GHz span power spectrum in 10-kHz RBW. (**C**) Detail of power spectrum.